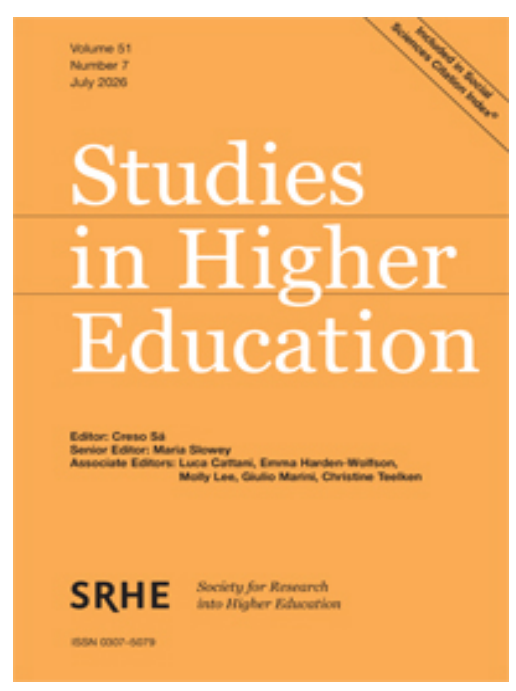

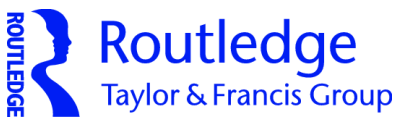

# A framework for developing university policies on generative AI governance: a cross-national comparative study

**Ming Li, Qin Xie, Ariunaa Enkhtur, Shuoyang Meng, Lilan Chen, Beverley Anne Yamamoto, Fei Cheng & Masayuki Murakami**

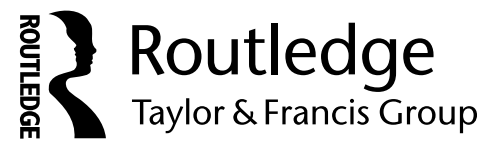

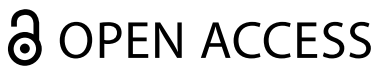

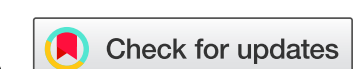

# A framework for developing university policies on generative AI governance: a cross-national comparative study

Ming Li [a], Qin Xie [b], Ariunaa Enkhtur [c], Shuoyang Meng [d], Lilan Chen [e], Beverley Anne Yamamoto [c], Fei Cheng [f,g] and Masayuki Murakami [h]

[a]Institute for Transdisciplinary Graduate Degree Programs, The University of Osaka, Osaka, Japan; [b]Comparative and International Development Education, University of Minnesota, Minneapolis, USA; [c]Institute for Global Initiatives, The University of Osaka, Osaka, Japan; [d]Institute for Advanced Studies for Asian Studies, The University of Tokyo, Tokyo, Japan; [e]Center for Higher Education Studies, Waseda University, Tokyo, Japan; [f]Research and Development Center for Large Language Models, National Institute of Informatics, Tokyo, Japan; [g]Graduate School of Informatics, Kyoto University, Kyoto, Japan; [h]Center for Education in Liberal Arts and Sciences, The University of Osaka, Osaka, Japan

**ABSTRACT**

As generative AI (GAI) becomes increasingly embedded in higher education, universities worldwide are developing policies to govern its ethical, pedagogical, and institutional use. However, these policies vary across national and institutional contexts. We undertake a cross-national analysis of GAI guidelines issued by leading universities in the United States, Japan, and China, identifying key policy orientations and proposing a structured framework to support policy development. Using an extended Technology Acceptance Model as an analytical lens, we examine five domains – Perceived Usefulness and Perceived Ease of Use, Perceived Risk, Facilitating Conditions, Social Influence, and Self-Efficacy, and identify 20 key themes through thematic coding. Together, these findings inform the development of the University Policy Development Framework for Generative AI (UPDF-GAI). Among the sampled institutions, U.S. universities emphasize faculty autonomy, practical application, and policy adaptability, reflecting environments shaped by cutting-edge research and peer collaboration. The Japanese universities analyzed adopt a more government-aligned approach, prioritizing ethics and risk management, but offering comparatively limited guidance on AI implementation and flexibility. The Chinese universities in the sample reflect a centralized, government-led model, focusing on technology application rather than early policy formulation, while actively exploring GAI integration in education and research. Based on these insights, the study proposes the UPDF-GAI, integrating technological, organizational, and social dimensions of policy formation. The framework provides a structured approach for universities to assess policy priorities, navigate tensions between innovation and risk, and strengthen institutional capacity for sustainable GAI governance, contributing to the evolving discourse on AI governance in higher education.

**CONTACT** Ming Li li.ming.itgp@osaka-u.ac.jp Institute for Transdisciplinary Graduate Degree Programs, The University of Osaka, 1-1 Machikaneyama, Toyonaka, Osaka 560-0043, Japan

## Introduction

Generative AI (GAI) technologies are becoming increasingly prevalent in higher education, transforming how universities approach teaching, learning, research, and administrative functions (Régis et al. 2024; UNESCO IESALC, 2023). In teaching, GAI supports the development of instructional materials (Baidoo-Anu and Owusu Ansah 2023; Kooli 2023a), assessment creation (Kooli and Yusuf 2024), and pedagogical adaptability for diverse student needs (Kooli and Chakraoui 2025; Latif et al. 2023). For learners, GAI provides personalized tutoring, language support, and immediate feedback, fostering independent learning (Farrokhnia et al. 2024; Kasneci et al. 2023). In research, it aids in data analysis, literature synthesis, and hypothesis formulation (Al-Zahrani 2024; Curtis 2023). Administratively, GAI facilitates routine inquiries, scheduling, and student services, enhancing operational efficiency and accessibility (Chan 2023; Li et al. 2024).

While GAI offers significant benefits to higher education, it also presents considerable challenges, particularly in academic environments that prioritize originality, integrity, and critical thinking (Sullivan, Kelly, and McLaughlan 2023; Tlili et al. 2023). Its application in academic writing and research grant proposals raises concerns regarding the authenticity of scholarly work and the attribution of intellectual contributions (da Silva 2023; Dwivedi et al. 2023). Moreover, excessive reliance on AI tools may undermine students' and researchers' critical thinking and academic competencies while blurring the distinction between independent work and AI-assisted content, thereby posing risks to academic integrity (Kaplan-Rakowski et al. 2023; Kooli and Chakraoui 2025; Ojha et al. 2023). Another critical issue is GAI's potential to reinforce biases embedded in training data, such as those related to gender and nationality, which may compromise the fairness and accuracy of educational outputs (Latif et al. 2023; Li et al. 2025). Additionally, the generation of spurious references by AI can contribute to misinformation, eroding trust in academic sources and scholarly discourse (Curtis 2023; Kooli 2023b). Given these challenges, it is imperative for higher education institutions to establish clear guidelines to ensure the responsible, transparent, and ethical integration of GAI in academic practices.

At the global level, organizations such as UNESCO and the European Union have issued guidelines to support AI governance in education, while national responses vary significantly across countries. By the end of 2023, only 18 of 38 OECD member countries had released provisional guidelines on GAI in education, revealing both shared concerns and divergent regulatory priorities. Differences are also evident in linguistic and cultural emphases, with non-English-speaking countries placing greater importance on cultural and linguistic adaptability (Vidal, Vincent-Lancrin, and Yun 2023). At the institutional level, universities have begun developing policies and training initiatives, yet substantial variation persists in their approaches to GAI governance (Nagpal 2024; Wang et al. 2024). For example, among the top 500 universities in the 2022 QS World University Rankings, fewer than one-third had established formal policies on ChatGPT, and among those, policy orientations differed considerably (Xiao, Chen, and Bao 2023).

These variations suggest that university GAI governance is shaped by multiple factors, including national regulatory environments, institutional cultures, and technological priorities. Initial studies indicate that U.S. universities often emphasize pedagogical innovation and ethical frameworks, whereas many Asian universities place greater emphasis on regulation, academic integrity, and institutional control (Dai et al. 2024; McDonald et al. 2024; Wu, Zhang, and Carroll 2024). Such diversity implies that relying on the experience of a single country or institution is insufficient to understand the complexity of GAI policy development. Rather than pursuing a universally applicable policy model, there is a need for a comparative and context-sensitive framework that identifies key policy domains and provides a structured reference for universities to design policies aligned with their specific institutional and national contexts.

To address this need, this study conducts a cross-national comparative analysis of university-level GAI policies in the United States, Japan, and China. Drawing on an extended Technology Acceptance Model (TAM; Davis 1989), the study examines five key domains – Perceived Usefulness (PU),

Perceived Ease of Use (PEOU), Perceived Risk (PR), Facilitating Conditons (FC), Social Influence (SI), and Self-Efficacy (SE) (Hameed and Arachchilage 2020; Jeyaraj, Rottman, and Lacity 2006; Venkatesh et al. 2003) – to analyze how universities conceptualize and govern the integration of GAI. Through qualitative content analysis of 124 policy documents from 110 universities, the study identifies recurring themes and patterns across different institutional and national contexts.

This study makes three main contributions. First, it advances the literature on GAI governance in higher education by providing a cross-national comparative perspective that integrates both Western and non-Western institutional contexts, addressing the current imbalance in existing research. Second, it extends the application of the TAM from individual technology adoption to institutional policy-making, offering a novel analytical lens for understanding how universities formulate AI governance strategies. Third, it develops the University Policy Development Framework for Generative AI (UPDF-GAI) as a structured yet flexible reference tool that enables universities to interpret, evaluate, and adapt GAI policies according to their specific institutional and national contexts. Rather than proposing a universal policy model, the framework supports context-sensitive policy development and helps institutions navigate the balance between innovation and risk in GAI governance.

Based on the five domains of the UPDF-GAI framework, we propose the following research questions for analysis the GAI guidelines of leading universities in the United States, Japan, and China.

RQ1. How do universities emphasize **PU** and **PEOU** in their GAI guidelines?

RQ2. How do universities address **PR** in formulation of GAI guidelines?

RQ3. What **FC** are provided to aid the adoption of GAI?

RQ4. How does **SI** shape the development of university GAI guidelines?

RQ5. How do university guidelines reflect institutions' **SE** in managing GAI?

## Literature review

### *International GAI policies for education*

UNESCO has played a key role in integrating GAI into education. In April 2023, it released *ChatGPT and Artificial Intelligence in Higher Education* (Sabzalieva and Valentini 2023), addressing challenges and ethical concerns while providing practical recommendations. This was followed by the *Guidance for Generative AI in Education and Research* (Miao and Holmes 2023) in September, emphasizing a human-centered approach, data privacy, ethical standards, and enhanced teacher training.

In September 2024, UNESCO introduced the *AI Competency Framework for Students* to help educators integrate AI learning into curricula and equip students with responsible and creative AI skills (Miao and Shiohira 2024). The framework focuses on four key areas: human-centered thinking, AI ethics, technical applications, and system design. These initiatives reflect UNESCO's commitment to ethical and inclusive AI in education.

Similarly, the European Union has promoted ethical AI use in education and research. In 2022, the European Commission issued guidelines to foster ethical AI practices and counter misconceptions (European Commission 2022; High-Level Expert Group on Artificial Intelligence 2022). The *2023 AI Report* highlighted AI's transformative role in teaching, learning, and administration, while underscoring ethical considerations, inclusivity, and innovation (European Commission 2023).

Building on these efforts, the European Research Area Forum published the *Living Guidelines on the Responsible Use of Generative AI in Research* in 2024, stressing transparency, accountability, and ethics (European Research Area Forum 2024). The guidelines urge researchers to disclose AI use, assess AI-generated content for bias, and avoid AI in sensitive tasks such as peer review. They also advocate for AI training in research institutions and fairness in AI-related funding decisions.

These international frameworks provide not only ethical principles and competency guidelines but also normative expectations that shape how universities develop their policies. By aligning with UNESCO and The European Union standards, such as human-centered approaches, transparency, and accountability, institutions seek global legitimacy and international collaboration. Understanding these policies is therefore crucial for analyzing the institutional guidelines examined in this study.

## Governmental GAI policies for education in the United States, Japan and China

As GAI advances, countries worldwide are developing distinct regulatory frameworks (Luna et al. 2024). In 2023, the United States, Japan, and China introduced policies reflecting their unique priorities in AI-driven education (Xie, Li, and Enkhtur 2024). This study selects the United States, Japan, and China based on their technological development, governance approaches, and AI policy strategies in higher education. The United States, as a leader in AI research and innovation, has a diverse and autonomous university policy system that fosters academia-industry collaboration (The Write House 2025). Japan emphasizes long-term AI planning, human-AI coexistence, and intelligent automation, with government guidance and significant university autonomy to ensure ethical advancement (Cabinet Office, Government of Japan 2022). China, leveraging its fast-growing AI industry and government-driven policies, adopts a top-down approach to university governance. Comparing these three nations highlights the impact of different policy structures on GAI governance and offers insights for universities developing globally aligned AI policies (Rehman et al. 2025).

The United States adopts a pragmatic approach in *Artificial Intelligence and the Future of Teaching and Learning*, emphasizing AI's role in enhancing efficiency, accessibility, and cost-effectiveness in K-12 education (U.S. Department of Education Office of Educational Technology 2023). It asserts that AI should support rather than replace teachers, maintaining a human-led model while addressing algorithmic bias through inclusive measures.

Japan prioritizes a human-centered, ethics-focused strategy in higher education. The Ministry of Education, Culture, Sports, Science, and Technology (MEXT) issued *About the Handling of Generative AI in Teaching at Universities and Technical Colleges*, advocating AI as a complement to traditional teaching while emphasizing transparency, ethics, and diverse assessments to uphold academic integrity (MEXT 2023).

China's approach is centralized, with the Ministry of Education collaborating on cross-departmental directives, such as the *Interim Measures for the Management of Generative Artificial Intelligence Services* (Cyberspace Administration of China 2023). These measures align AI use with national security and innovation priorities, promoting AI integration in education while providing limited classroom-specific guidance, focusing instead on broader societal objectives.

Although framed at the national level, these policies directly shape university governance. In the United States, federal emphasis on accessibility and inclusion informs GAI curricula; in Japan, MEXT embeds ethics and transparency in pedagogy; and in China, centralized directives link institutional adoption to national priorities. Examining these policies thus provides a necessary backdrop for understanding institutional responses.

## University level of GAI policies and guidelines

University-level GAI guidelines emerge within this broader international and national policy environment. While institutions develop their own governance strategies, they are both enabled and constrained by global frameworks and national regulations. Against this backdrop, research has increasingly focused on its governance in universities. An empirical study of QS (Quacquarelli Symonds) Top 500 universities suggests that GAI policy formulation is influenced by academic reputation, English-speaking status, and public attitudes toward AI (Xiao, Chen, and Bao 2023). Similarly, Dabis and Csáki (2024) examined Academic Ranking of World Universities (ARWU) Top 500 universities, finding that most require students to demonstrate personal knowledge in assignments and take responsibility for AI misuse. Additionally, a study of the world's top 50 universities highlights the role of academic integrity, assessment design, and student communication in policy development (Moorhouse, Yeo, and Wan 2023).

At the national level, McDonald et al. (2024) found that over half of U.S. R1 universities encourage GAI use in teaching and offer model syllabi for curriculum design. Wu, Zhang, and Carroll (2024) analyzed AI governance in Big Ten universities and proposed three models: multi-departmental collaboration, role-based guidance, and academic-driven approaches for responsible AI use.

In Asia, Dai et al. (2024) examined QS Top 60 Asian universities, emphasizing academic integrity and text generation guidelines but identified gaps such as limited stakeholder engagement, weak empirical support, misalignment with global standards, and reliance on traditional academic paradigms.

Current research mainly focuses on Western universities and global rankings, often overlooking challenges in non-Western institutions or cross-national comparisons. Although some studies include Asian universities, they treat them as a uniform group, lacking regional specificity. Moreover, existing research falls short of developing a comprehensive framework for global GAI policy advancement (Batista, Mesquita, and Carnaz 2024).

To address these gaps, this study selects leading universities in the United States, Japan, and China, integrating Western and non-Western perspectives while incorporating local ranking systems to better reflect regional contexts. It further proposes a systematic framework to support coordinated GAI policy development and provide actionable guidance for higher education institutions.

## Methodology

This study employs qualitative content analysis, a systematic and flexible approach for interpreting qualitative data (Schreier 2012). Content analysis is widely recognized for its versatility and adaptability (Schreier et al. 2019), particularly in the analysis of multilingual and multicultural materials (Bethmann and Niermann 2015).

Comparative education has long relied on analytical frameworks rooted in Western perspectives (Hayhoe et al. 2017). Many university comparative studies use global rankings as selection criteria; however, these ranking systems, such as QS World University Rankings, Times Higher Education World University Rankings, and ARWU, exhibit a notable bias toward Western institutions (Kochetkov 2024).

According to 2024 data, the top 30 universities across these three ranking systems include 91 institutions, because in the QS and Times Higher Education rankings there are ties, which means that more than 30 universities are included within the top 30 positions. North American and European universities dominate, accounting for 84.6% (United States: 51, United Kingdom: 13, Canada: 4, France: 2, Germany: 1, Switzerland: 3, Australia: 3). In contrast, Asian universities comprise only 15.4% (China: 7, Japan: 3, Singapore: 3, Hong Kong: 1), while African and Latin American universities are entirely absent.

This imbalance highlights the Western-centric bias embedded in global university rankings. As Welsh (2021) asserts, 'Although ranking systems may appear to evolve in form, their logic, structure, and impact remain deeply political' (28). Overreliance on these rankings in research selection processes can lead to analytical bias, systematically undervaluing non-Western institutions.

To mitigate such bias, this study deliberately includes both Western and non-Western universities, ensuring a more comprehensive analysis of GAI guideline development across diverse educational and cultural contexts. Local ranking systems, which reflect country-specific evaluation criteria, are employed instead of global rankings. Furthermore, rather than selecting a single university to represent an entire nation (Tobin, Hsueh, and Karasawa 2009), multiple institutions are considered to enhance sample diversity and generate broader insights.

## Data collection

We reference prominent local ranking systems to identify leading universities in each country. For the United States, we consider the top 30 institutions from the *U.S. News Best Colleges 2024* list (31 universities in total), recognized for their strong emphasis on academic freedom and technological innovation (Webster 2001). In Japan, our study focuses on the 37 universities selected for the *Top Global Universities Project,* an initiative funded by MEXT to drive comprehensive university reform and internationalization between 2014 and 2023 (Black 2022). In China, our analysis centers on the *Double First-Class* universities, distinguished as premier institutions leading national research and innovation efforts (Huang 2015).

We focus exclusively on GAI policies and guidelines officially released at the university level from 2023 to November 2024, treating these as our primary data sources. Policies and guidelines issued by individual departments, programs, or faculty members are intentionally excluded to maintain consistency and comparability in the data set.

We then conducted a systematic search of the official websites of these institutions, using keywords such as 'AI,' 'GAI,' 'guideline,' 'policies,' and 'ChatGPT' in English, Japanese, or Chinese, depending on the country, to locate relevant GAI policies and guidelines. By focusing only on policies and guidelines sanctioned at the university level, we ensure a level of authority and standardization across the data, which enables a more accurate and meaningful comparison of institutional approaches to AI technology adoption and governance.

## Framework development

The TAM proposed by Davis (1989), was initially designed to examine individuals' acceptance and adoption of new technologies. Over time, its application has expanded to organizational contexts (Hameed and Arachchilage 2020). In universities, technology-related policies are shaped primarily by administrators whose perceptions of emerging technologies influence institutional strategies and guidelines.

While the TAM provides a strong theoretical foundation, its applicability is limited in environments where technological, ethical, social, and practical factors intersect. To better explain university administrators' technology acceptance behavior, this study extends the core TAM constructs – PU and PEOU – by incorporating PR, FC, SI, and SE. Together, these constructs form the UPDF-GAI framework, which reflects both technological evaluations and broader institutional conditions (Figure 1).

Although the TAM encompasses the constructs Attitude Toward Usage, Behavioral Intention to Use, and Actual Usage (Davis 1989; Venkatesh et al. 2003), these dimensions are not incorporated in the study. This is because the analytical focus is on institutional policy formulation rather than individual technology adoption. At the university level, administrators' assessments of usefulness, feasibility, risks, resources, and social pressures are more directly related to policy development than individual users' behavioral intentions.

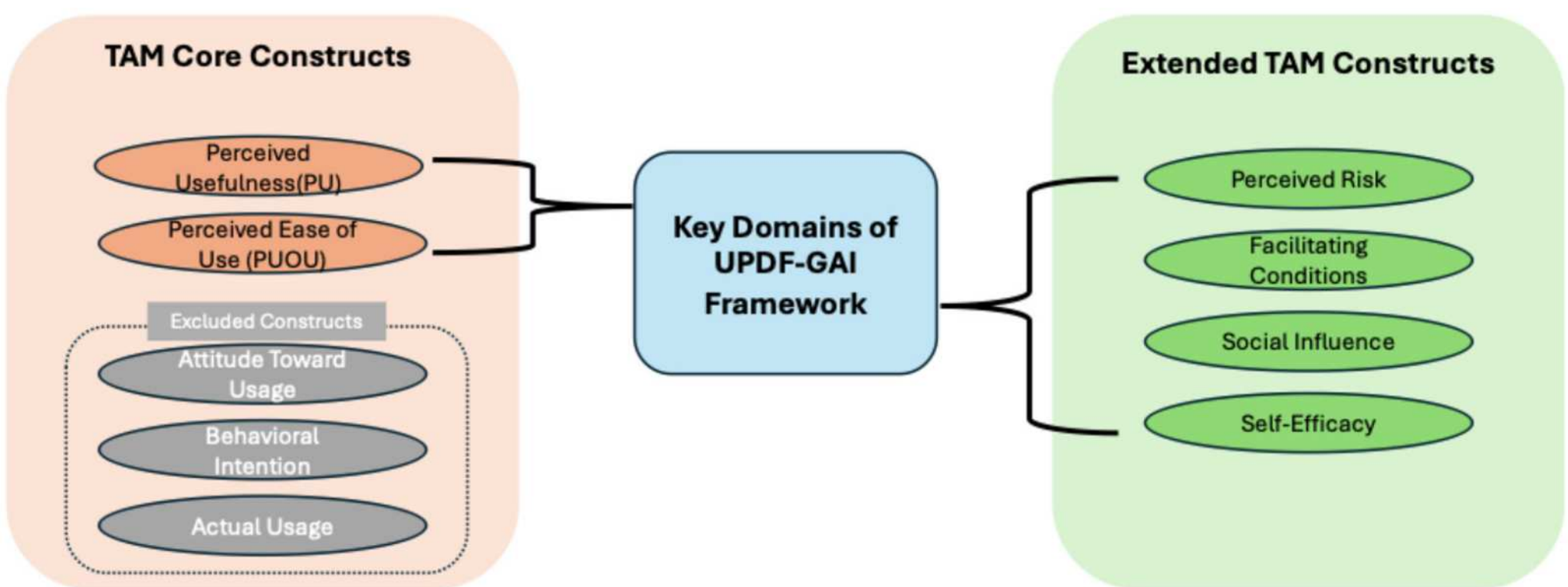

**Figure 1.** Key Domains of UPDF-GAI Framework.

The inclusion of SI, FC, PR, and SE is supported by prior research on organizational technology adoption (Hameed and Arachchilage 2020; Jeyaraj, Rottman, and Lacity 2006) and by the specific context of generative AI in higher education. These constructs capture key organizational and social dimensions, such as institutional support, ethical concerns, peer expectations, and decision-makers' confidence, that shape how universities design and implement GAI policies.

### *PU and PEOU*

PU and PEOU are core constructs of TAM, explaining users' technology adoption behavior (Davis 1989). PU refers to the perceived benefits of a technology, such as improved performance and efficiency, while PEOU reflects its ease of use, including intuitive design and operational simplicity.

In the context of university policies on GAI, PU captures how institutions evaluate GAI's potential to enhance learning outcomes, drive research innovation, and improve administrative efficiency. PEOU, on the other hand, reflects policy measures that simplify operational processes, provide clear guidelines, and enhance accessibility to reduce perceived barriers, thereby increasing GAI adoption. This study integrates PU and PEOU into a single core domain, as policymakers typically evaluate both the value and feasibility of a technology to ensure its effective implementation without significant adoption barriers.

### *PR*

PR refers to potential negative consequences associated with technology use (Featherman and Pavlou 2003; Martins, Oliveira, and Popovič 2014). In the context of GAI, this includes concerns about privacy, data security, ethical use, and biases within AI systems. For universities, addressing these risks in GAI guidelines is essential to ensure that users feel confident in employing AI tools without compromising ethical standards or academic integrity. PR is a critical construct for analyzing university policies because it sheds light on how institutions anticipate and manage the potential downsides of AI adoption. By acknowledging and addressing these risks, universities seek to build trust in AI technologies and encourage their use in ways that align with both institutional priorities and societal values.

In our analysis, we examine how GAI guidelines from universities across different countries address PRs, particularly in terms of privacy, ethics, and security. This approach helps us understand how institutions in various cultural and political contexts mitigate the challenges associated with AI use.

### *FC*

FC refer to the infrastructure, resources, and support systems that enable effective technology use (Holden and Karsh 2010; Lu et al. 2003). In an educational context, this includes access to training programs, technical support, and institutional resources that help faculty, staff, and students use GAI tools effectively. FC are crucial for building user confidence and ensuring that AI adoption is not impeded by practical barriers. By offering adequate support, institutions can encourage smoother and more effective integration of AI tools into academic activities.

In our analysis, we explore how universities in the United States, Japan, and China incorporate FC into their AI guidelines. We examine whether these policies provide sufficient support, such as training programs and technical assistance, and assess how these factors impact the ease of GAI adoption across different institutions.

### *SI*

SI reflects the extent to which individuals' technology adoption is shaped by expectations from peers, institutional leadership, and regulatory bodies (Hsu and Lin 2008; Venkatesh et al. 2003; Venkatesh and Morris 2000). In educational settings, this can manifest through institutional expectations, peer pressure, or compliance with external policies like government regulations. For universities, SI plays a key role in shaping GAI guidelines, often guided by national strategies, international trends, or the policies of peer institutions. Some universities may develop GAI guidelines in alignment with broader external directives, while others may focus on internal priorities driven by faculty, staff, or student demand, reflecting less reliance on external pressures.

Analyzing SI helps us understand how both external and internal factors shape AI policies. This approach allows us to assess whether universities are primarily reactive to external influences or whether their guidelines reflect more internally driven priorities.

### *SE*

SE refers to an individual's confidence in successfully using a technology (Compeau and Higgins 1995; Holden and Rada 2011; Mun and Hwang 2003). In the context of GAI policy development, it reflects the confidence universities have in creating clear, actionable guidelines. High SE is demonstrated by well-developed, detailed policies that include clear protocols and ethical guidelines, while low SE may result in vague or incomplete policies, reliance on external frameworks, or hesitation to take responsibility for GAI adoption.

Our analysis examines how universities exhibit SE through their GAI guidelines, assessing whether institutions develop robust, research-backed policies or defer to external sources. For instance, some universities may release preliminary drafts or no guidelines at all, suggesting lower SE in managing GAI. By exploring SE in policy development, we can better understand how universities perceive their ability to regulate and implement GAI, providing insight into the role SE plays in shaping effective and responsible GAI adoption.

## *Data analysis*

Data analysis begins with background coding, extracting key contextual information from GAI policies and guidelines, including issuing institutions, publication dates, languages, and other relevant metadata to establish a standardized basis for comparison. Building on this, we primarily adopted a deductive, theory-driven approach to thematic analysis. Specifically, the analysis was guided by the five domains of the UPDF-GAI framework: PU and PEOU, PR, FC, SI, and SE. Relevant content in the guidelines was first identified and coded into the corresponding domains.

Because the guideline texts were often lengthy and diverse, and many passages conveyed similar meanings, the initial coding process involved condensing and abstracting the textual material into descriptive codes that captured the central meaning of the guideline statements. For example, statements related to information protection were grouped under the code 'data security,' while those referring to ownership of content were coded as 'copyright infringement.' These descriptive codes then served as the basis for further synthesis within each UPDF-GAI domain, which ultimately led to the identification of higher-level themes. For instance, within the PR domain, codes such as 'data security' and 'copyright infringement' were consolidated into the theme 'security and intellectual property risks,' while codes such as 'plagiarism,' 'false or inaccurate information,' and 'legal risks' were synthesized into the theme 'ethical and legal risks.'

It should be noted that some categories (e.g. data security and copyright infringement) may conceptually overlap with ethical concerns. In this study, these issues are categorized based on their primary policy focus in institutional guidelines, where security-related concerns are distinguished from broader ethical and legal considerations to enhance analytical clarity. This analytic strategy ensured that the themes remained consistent with the theoretical framework while also retaining sensitivity to the nuances present in the policy texts. Although thematic coding is less automated than computational approaches, it enables more precise interpretation of policy meanings and facilitates the identification of nuanced themes and sub-themes that cannot be readily captured by algorithmic methods.

To minimize researcher bias and confirmation bias, multiple researchers were involved in the coding process. The policy documents were originally written in three languages – Chinese, Japanese, and English – so coding responsibilities were allocated based on linguistic proficiency and contextual knowledge of national higher education systems to ensure accurate interpretation of policy content. Accordingly, the first author analyzed policies from China and Japan, while the second author analyzed U.S. policies. At the initial stage, both authors independently coded a subset of the data, followed by discussions to reach consensus on how specific codes should be grouped into themes. After completing the full coding, the themes were revisited and jointly refined to ensure consistency across coders. This iterative process of independent coding and collaborative discussion enhanced the transparency and credibility of the analysis.

## Findings

### *Overview*

In the United States, 31 universities were included in the analysis, all of which have issued at least one guideline on the use of GAI, resulting in a total of 80 guidelines. 26 universities have released multiple guidelines each, reflecting a proactive and comprehensive approach to GAI policy development.

In Japan, among 37 designated *Top Global Universities Project* universities, a total of 37 guidelines were identified. Notably, 10 universities have not issued any guidelines, while 8 institutions have published more than one, indicating uneven adoption of GAI policies across universities.

The situation in China presents a marked contrast. Among 42 *Double First-Class* universities, only 6 institutions have released a total of 7 guidelines, suggesting limited engagement with GAI policy formulation at the institutional level.

Universities in the United States and Japan began releasing guidelines on GAI as early as the beginning of 2023, following the launch of ChatGPT. In contrast, only one university in China issued guidelines in 2023, with the remainder releasing theirs in 2024. This discrepancy suggests that while universities in the U.S. and Japan responded promptly to the rapid advancements and implications of GAI, Chinese universities exhibited a slower response. This may be attributed to differences in policy-making processes, technological readiness, or risk assessment strategies.

### *RQ1: how do universities emphasize PU and PEOU in their GAI guidelines?*

An analysis of target universities in the United States, Japan, and China reveals that none explicitly prohibit the use of GAI. Rather, most institutions advocate for its cautious and responsible use within an ethical framework, emphasizing ethical considerations. However, significant differences exist among these countries in terms of the specificity and proactiveness of institutional guidelines.

To examine PU and PEOU of GAI policy in higher education, this study conducted a thematic analysis of university policies and categorized GAI applications into four themes: (1) **Content Generation and Editing**, encompassing the creation and refinement of digital content such as images, videos, articles, and code; (2) **Personalization and Assistive Functions**, including language translation and real-time communication support; (3) **Efficiency Enhancement and Automation**, which involves data analysis, minor code debugging, and information retrieval; and (4) **Support for Creativity and Innovation**, wherein GAI facilitates brainstorming and collaborative discussions. These four themes are incorporated into the UPDF-GAI framework. Figure 2 shows the proportion of universities in these three countries that have published guidelines covering these themes.

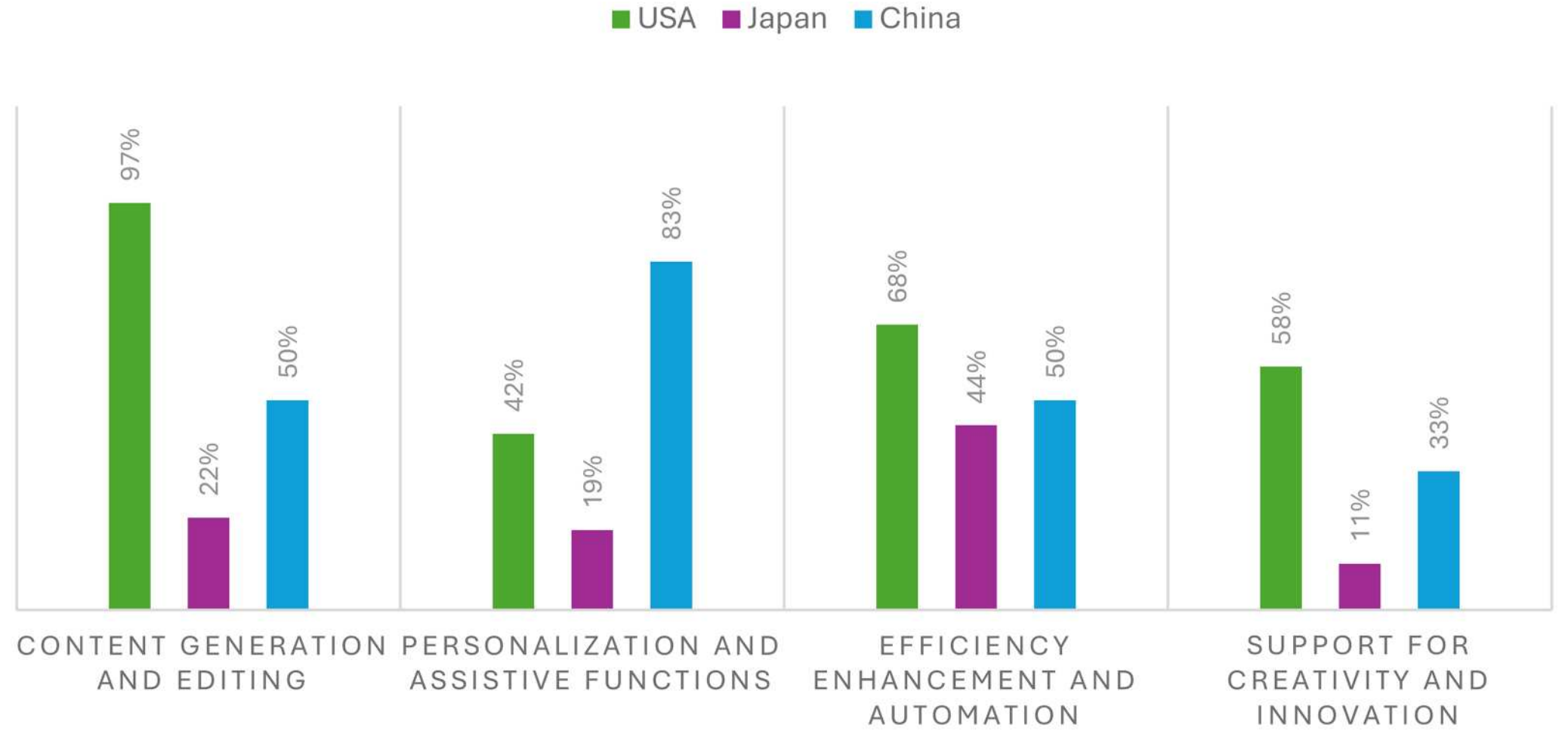

**Figure 2.** Percentage of Universities that Address PU and PEOU in Their Guidelines by Country, sample size by country: USA (n = 31), Japan (n = 27), China (n = 6).

### *United States: emphasizing responsible use and innovation*

U.S. universities acknowledge the potential of GAI while underscoring the importance of responsible usage. Specifically, the California Institute of Technology (n.d.) encourages students to integrate GAI into their academic practices as part of information literacy. Brown University (2023, August 31) promotes experimental engagement with GAI to enhance learning experiences, while Harvard University (Garber, Weenick, and Jelinkova 2023) and Yale University (Strobel, Callahan, and Barden 2023) strongly advocate for its incorporation into education. Notably, 97% surveyed U.S. universities emphasize GAI's role in Content Generation and Editing, often providing concrete examples of its application in personalized learning. Additionally, these institutions recognize GAI's ability to improve research and administrative efficiency and foster creativity and innovation among faculty and students.

### *Japan: cautious adoption with limited practical guidance*

In Japan, universities such as the University of Tokyo (2023, May 26), University of Tsukuba (2023, May 11), and Nagoya University (2023, July 5) explicitly endorse the effective use of GAI tools, whereas others, including Tohoku University (2023, April 24) and Science Tokyo (2023, April 20), acknowledge the impracticality of an outright ban. However, compared to the United States and China, Japanese university guidelines tend to be broader in scope and provide fewer practical examples. While some institutions, such as the University of Tokyo (2023, April 28) and The University of Osaka (n.d.), have offered specific applications of GAI in academic and administrative contexts, overall, Japanese universities exhibit a lower frequency of references to PU and PEOU. Discussions primarily focus on Efficiency Enhancement and Automation (44%), yet remain largely abstract, with limited elaboration on concrete use cases. References to creativity and innovation are even scarcer (11%), reflecting a generally cautious stance.

### *China: integrating GAI literacy into long-Term educational strategies*

Although fewer universities in China have published explicit GAI guidelines, efforts to integrate GAI literacy into teaching and research are evident. For example, Sichuan University (n.d.) and China Agricultural University (2024, November 15) provide foundational knowledge on GAI alongside its applications in academic settings. Notably, Zhejiang University (2024) has introduced the Red Book on AI Literacy for University Students, outlining a strategic framework for AI competency development. Chinese universities adopt a moderate approach toward GAI's role in content generation, with some institutions offering illustrative examples in educational contexts. Many universities emphasize its potential for language acquisition and communication skills, while some provide detailed guidelines on its use in research optimization and teaching refinement. Furthermore, several institutions highlight the importance of integrating GAI's innovative potential into broader educational objectives, aligning with national strategies to enhance students' overall competitiveness.

### *RQ2: how do universities address PR in formulation of GAI guidelines?*

Universities in the United States, Japan, and China have examined the ethical, legal, and social implications (ELSI) of GAI in considerable depth within their institutional guidelines. Through coding analysis of these policies, eleven specific ELSI concerns were identified: plagiarism, false or inaccurate information, data security, copyright infringement, legal risks, control of speech and thought, bias and discrimination, over-reliance on AI, AI divide, environmental risks, and educational risks. The cross-national distribution of these concerns is presented in Figure 3. Across all three nations, universities consistently emphasize issues such as data security, copyright infringement, plagiarism, and the dissemination of misinformation, which are widely recognized as fundamental challenges in the integration of GAI into higher education. Building on this analysis, the eleven concerns were further synthesized into six overarching themes: (1) **Security and Intellectual Property Risks**, (2) **Ethical and Legal Risks**, (3) **Social and Cultural Risks**, (4) **Psychological Risks**, (5) **Environmental Risks**, and (6) **Educational Risks**. These six themes are incorporated into the UPDE-GAI framework.

#### *United States: comprehensive consideration of ELSI issues*

Universities in the United States exhibit the broadest scope of ELSI considerations. Over 90% universities place strong emphasis on data security, particularly the need to protect sensitive information exchanged during AI input and output processes. Additionally, there is a heightened awareness of bias and discrimination (74%), reflecting a commitment to ensuring social justice in AI applications. However, despite this extensive coverage, U.S. universities rarely provide explicit discussions on other potential risks, such as over-reliance on AI, influence on discourse and cognition, or educational risks.

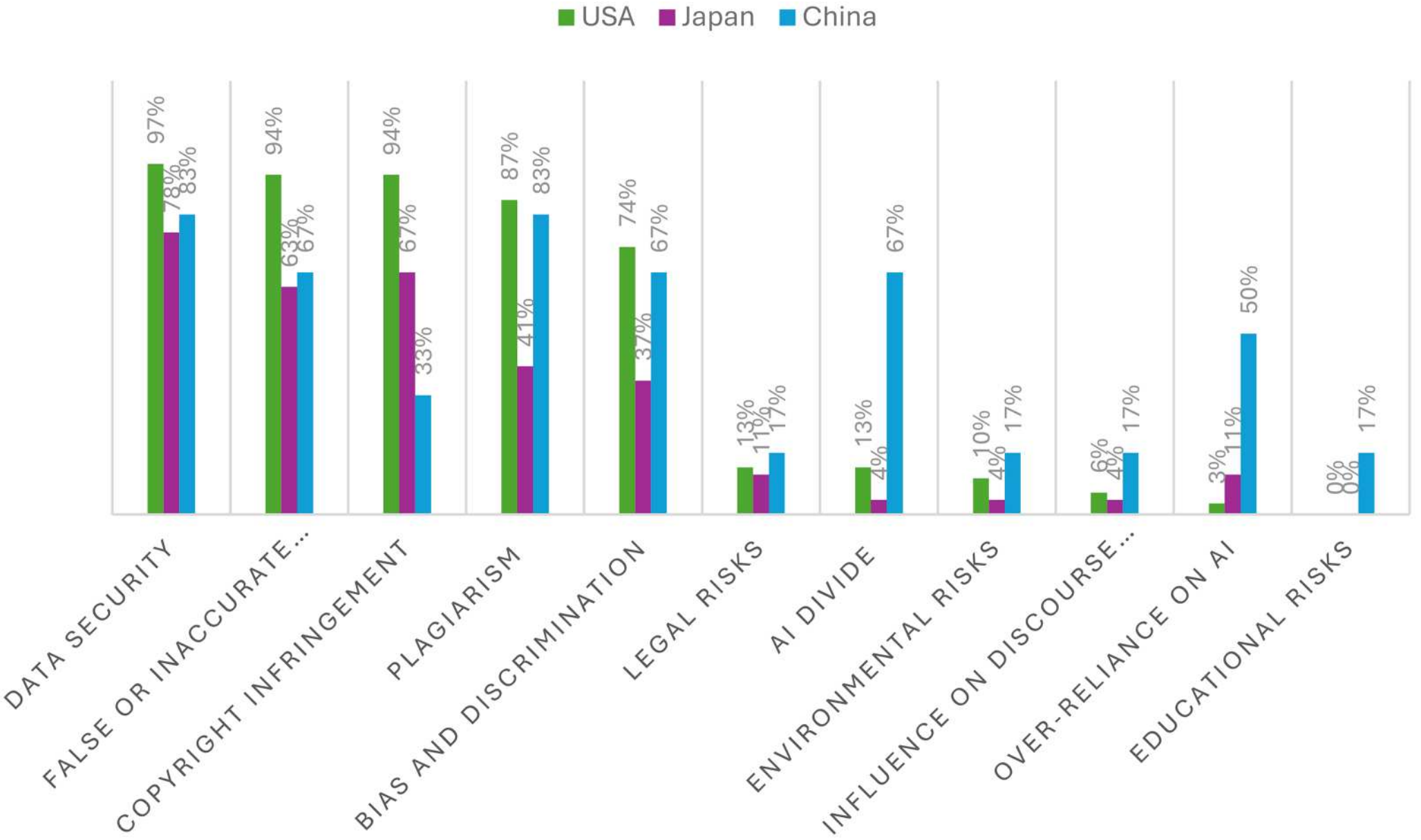

**Figure 3.** Percentage of Universities that Address PR by Country, sample size by country: USA (n = 31), Japan (n = 27), China (n = 6).

### *Japan: focus on security and copyright, limited discussion on SR*

The ELSI concerns addressed by Japanese universities largely align with those in the United States. For instance, Tohoku University explicitly acknowledge risks related to plagiarism and the spread of misinformation (2023, April 24; 2023, August 30). However, Japanese universities primarily focus on Data Security (78%) and Copyright Infringement (67%) with comparatively limited discussions on social and cultural risks. Furthermore, Japanese institutional guidelines rarely address topics such as the AI divide, influence on discourse and cognition, environmental risks, or educational risks, indicating that these issues remain underexplored within Japan's higher education sector.

### *China: diverse range of ELSI topics, emphasizing AI dependency and educational risks*

Although fewer Chinese universities have published explicit GAI guidelines, those that have done so cover a comparatively diverse range of ELSI concerns. In addition to the commonly discussed issues of data security, copyright, and plagiarism, Chinese universities also highlight the risks of over-reliance on AI and the AI divide. Notably, Zhejiang University (2024) explicitly raises concerns regarding educational risks, cautioning that the role of educators may be diminished and that students' excessive reliance on AI could lead to increasingly isolated learning experiences.

## *RQ3: what FC are provided to aid the adoption of GAI?*

Universities provide varying levels of support for AI adoption in educational settings. We have categorized references to FC in university guidelines into three themes: **Epistemological Support**, **Practical Support**, and **Policy Support** (Figure 4). These three themes are embedded within the UPDF-GAI framework.

Epistemological support helps faculty, students, and staff understand, acquire, and evaluate knowledge about GAI, fostering critical thinking and reflective application in teaching and learning. Practical support includes tangible resources such as training sessions, software access, and instructional materials. Policy support refers to GAI guidelines issued by universities or external organizations such as UNESCO and OECD.

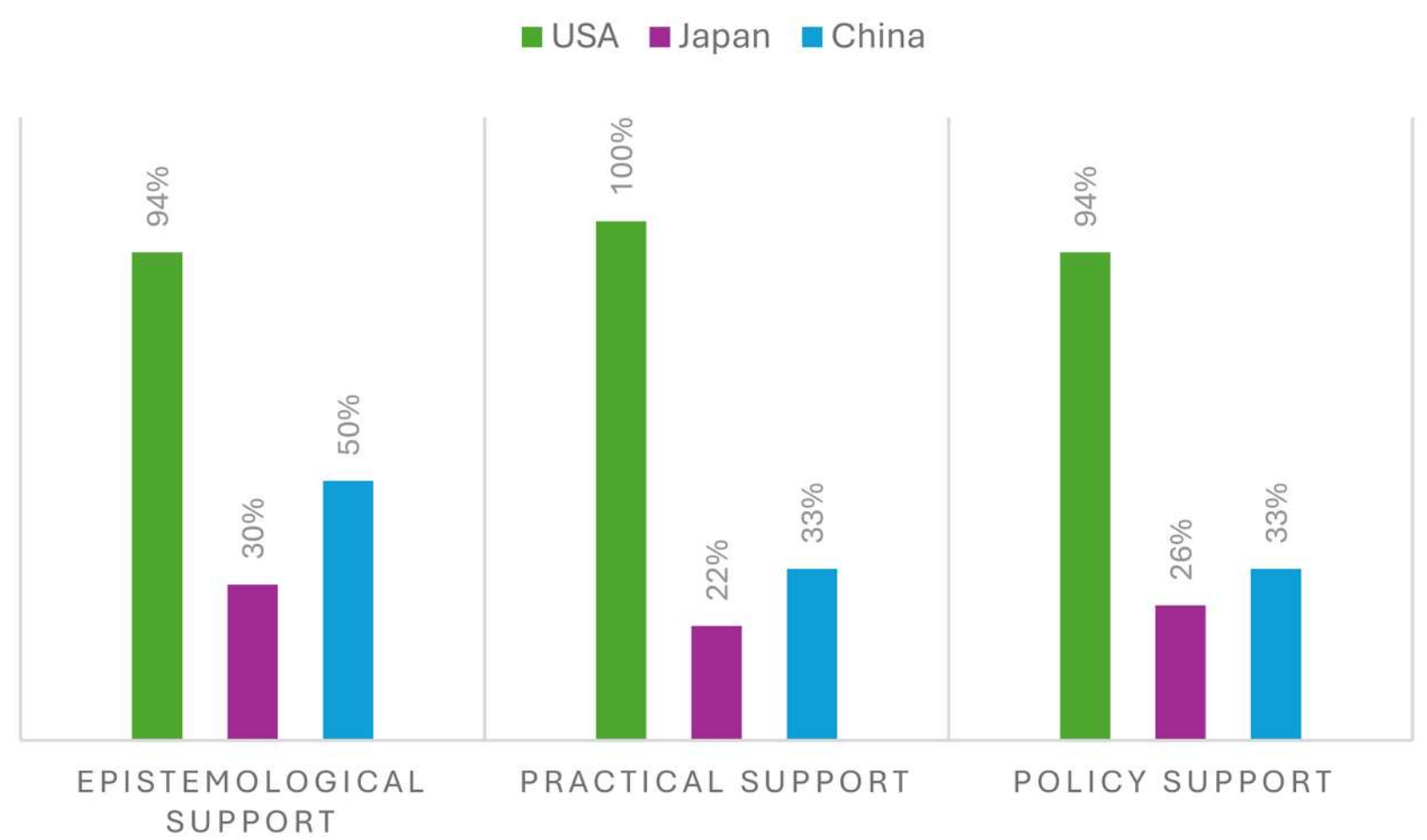

**Figure 4.** Percentage of Universities that Address FC by Country, sample size by country: USA (n = 31), Japan (n = 27), China (n = 6).

### *United States: comprehensive and balanced support*

In the United States, universities demonstrate comprehensive and balanced support across all three categories, with very high coverage for epistemological (94%), practical (100%), and policy support (94%). This indicates a robust approach to integrating GAI into academic settings. For example, Columbia University (n.d.) emphasizes the reflective use of GAI in educational practices. Yale University (Strobel, Callahan, and Barden 2023) categorizes and provides detailed information about AI tools, stating, 'These tools are provided to all Yale community members and are an excellent entry point for exploring GAI.' The university also includes clear descriptions of tools, access links, and data classification levels. Similarly, Harvard University (n.d.) incorporates comprehensive resources by attaching its information security policy, copyright policy, and school-based resources and policies to guide the ethical and secure use of GAI.

### *Japan: limited but emerging support measures*

Although fewer Japanese universities provide GAI support, institutions have implemented comprehensive measures to offer GAI knowledge guidance, practical support, and policy frameworks for both faculty and students. For example, the University of Tokyo (2023, April 28) encourages faculty members to evaluate the applicability of AI tools in various educational scenarios before integrating them into teaching and assessment practices. Tohoku University (2023, Augst 30) offers practical support by recommending strategies to faculty for mitigating potential misuse of AI in student assignments. Additionally, the university provides students with practical advice on using AI for research and academic writing. Furthermore, Sophia University (2023, October 23) and Hiroshima University (2023, May 23) have introduced faculty development programs to enhance educators' knowledge of GAI.

### *China: policy-driven approach with selective initiatives*

While Chinese universities offering GAI support are limited, certain institutions have taken significant steps to provide robust policy guidance. For instance, Sichuan University (n.d.) has issued a set of regulations titled 'Principles and Guidelines', supported by both domestic and international standards. Shanghai Jiao Tong University (2023) has improved teaching effectiveness by offering training programs for faculty and leveraging feedback generated by GAI to optimize instructional practices.

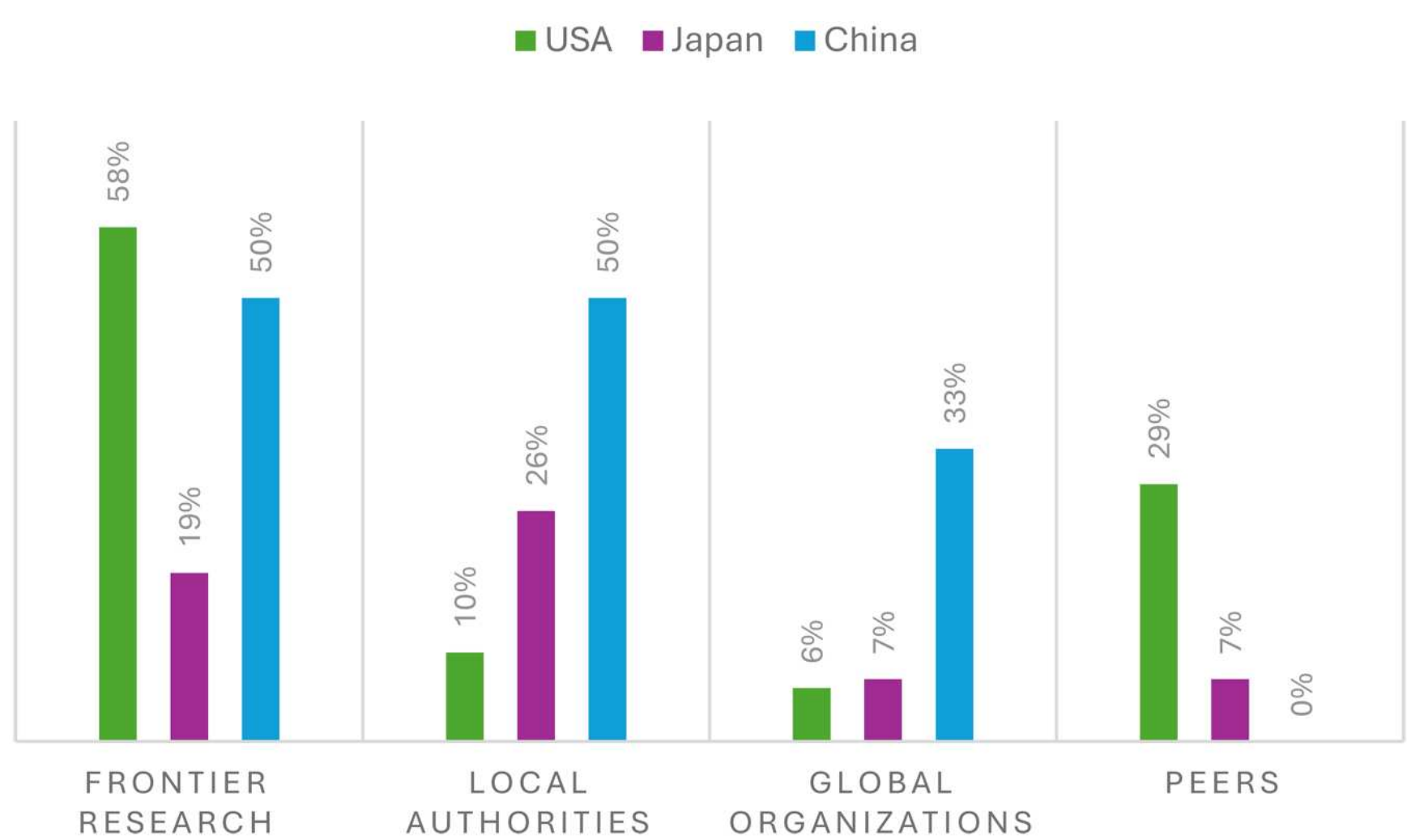

**Figure 5.** Percentage of Universities that Address SI by Country, sample size by country: USA (n = 31), Japan (n = 27), China (n = 6).

### *RQ4: how does SI shape the development of university GAI guidelines?*

Through thematic analysis, we summarize the social impact of generative AI into four themes: **Frontier Research**, **Local Authorities**, **Global Organizations**, and **Peers** (Figure 5). These themes are incorporated under the UPDF-GAI framework. Frontier Research refers to the latest information on GAI research, particularly in the context of higher education. Local Authorities denotes local governments, governmental leaders, or relevant governmental departments. Global Organizations includes entities operating across multiple nations and addressing issues that transcend national borders, such as UNESCO. Peers refer to post-secondary educational institutions and universities.

#### *United States: research-driven policies and strong peer collaboration*

U.S. universities prioritize Frontier Research (58%), demonstrating a reliance on academic and technological advancements to shape GAI policies. For example, Northwestern University (n.d.) curates the latest studies on GAI in education, while UCLA (Bawn and Bisley 2023) compiles media reports, such as those from The New York Times, to assess ChatGPT's impact on higher education. This highlights a strong innovation-driven approach, with minimal reliance on local authorities or global organizations. Additionally, peer collaboration plays a crucial role (29%), reflecting a robust academic and industry network that actively contributes to GAI policy development.

#### *Japan: government-driven policies with limited research influence*

Japanese universities exhibit moderate engagement with social impact factors, with government authorities, particularly MEXT, playing a dominant role in shaping GAI policies. 26% of university guidelines reference government policies (e.g. University of Tokyo, University of Osaka, Tohoku University). While frontier research holds some influence, decision-making is largely government-led. Universities, such as Tohoku University (2023, August 30), incorporate domestic and international case studies as additional references for policy development.

#### *China: government-led framework with academic support*

In China, Local Authorities (50%) and Frontier Research (50%) are the primary influences on GAI policies, reflecting a top-down, government-driven approach. National and local governments dictate policy direction, with research institutions contributing to technical guidelines and innovation. However, academic influence remains secondary to government directives. Unlike in the U.S., peer collaboration is minimal, indicating a centralized decision-making structure where policies are formulated within governmental and institutional frameworks rather than through broad stakeholder engagement.

### *RQ5: how do university guidelines reflect institutions' SE in managing GAI?*

Through thematic analysis, we identify three themes that reflect SE in university guidelines: **Instructors as Gatekeepers**, **Stakeholders' Feedback**, and **Adaptive Policy Frameworks** (Figure 6). Instructors as Gatekeepers refer to universities delegating the authority to instructors to decide whether to permit or restrict the use of GAI in educational settings. Stakeholders' Feedback involves universities actively seeking input from faculty, students, and staff to refine and adjust their policy-making processes. Adaptive Policy Frameworks reflect the flexibility of GAI policies, acknowledging that they may be provisional and subject to revision as technology evolves. Together, these themes illustrate varying levels of SE in how universities approach GAI integration. The three themes are incorporated into the UPDF-GAI framework.

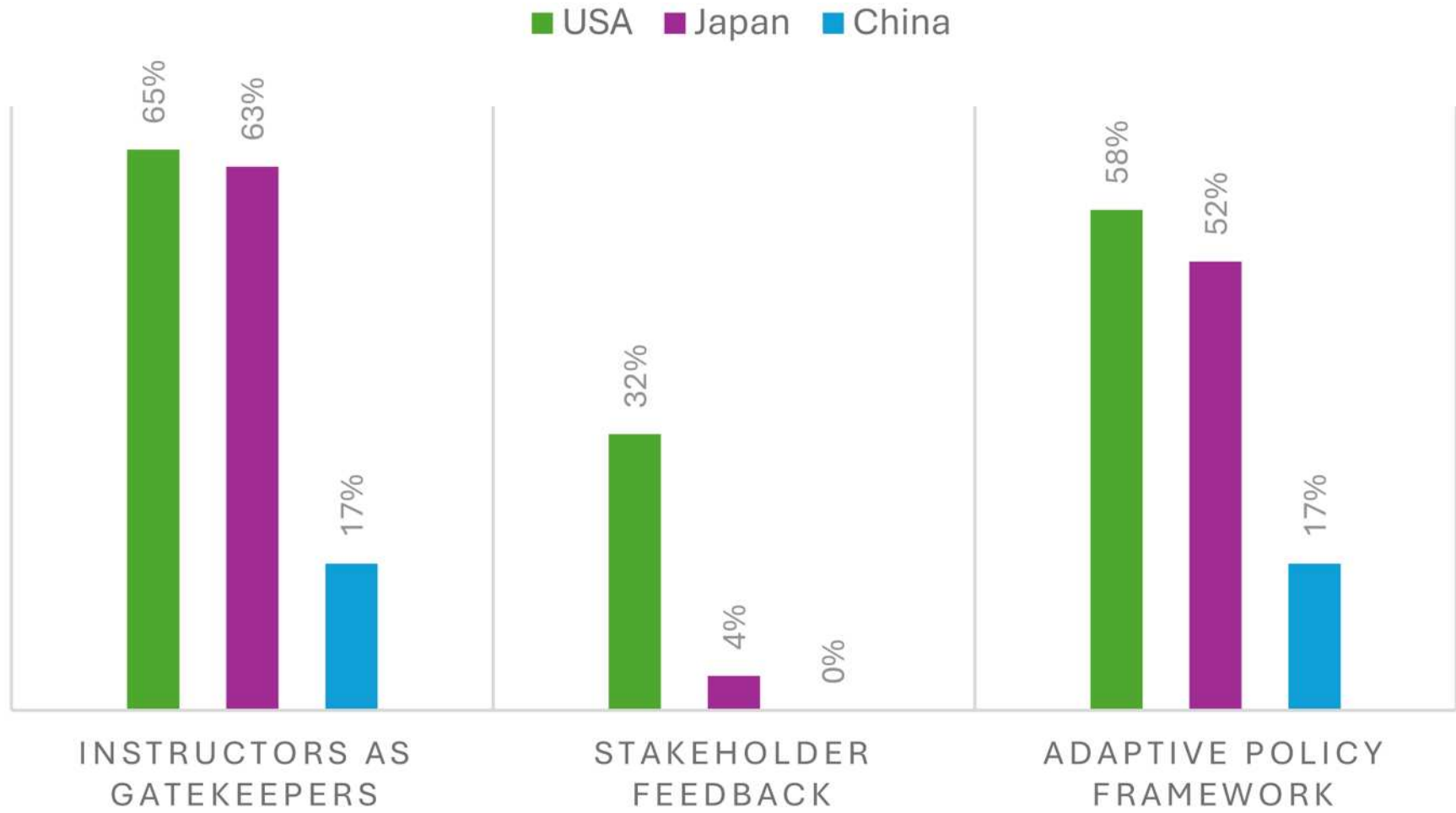

**Figure 6.** Percentage of Universities that Address SE by Country, sample size by country: USA (n = 31), Japan (n = 27), China (n = 6).

### *United States: strong faculty autonomy and policy flexibility*

65% of U.S. universities allow faculty to determine GAI use in their courses, emphasizing academic freedom and professional discretion. This decentralized approach enables instructors to adapt GAI integration to their specific disciplines and student needs. Additionally, 58% of universities adopt adaptive policy frameworks, recognizing the need for flexible regulations that evolve with technological advancements and educational demands. However, stakeholder feedback is less emphasized, with only 32% of universities incorporating faculty, student, or administrative input in policymaking. This suggests a preference for faculty autonomy over broad stakeholder consensus, reinforcing a governance model where academic professionals play a central role in decision-making.

### *Japan: moderately centralized governance with limited adaptability*

Japanese universities exhibit moderate reliance on faculty autonomy and adaptive policies, with 63% of institutions delegating some decision-making authority to instructors and incorporating policy flexibility. While faculty autonomy is acknowledged, national guidance situates responsibility primarily at the institutional level. For example, MEXT instruct universities and technical colleges to establish transparent institutional policies and assessment mechanisms regarding GAI use, thereby placing primary authority in administrative and managerial bodies rather than in individual educators (MEXT, 2023). Stakeholder feedback is also notably limited (4%), with minimal involvement from faculty and students in shaping institutional guidelines. This reflects a governance model where administrative oversight remains dominant, even when policies formally recognize some degree of faculty autonomy.

### *China: highly centralized policy framework with limited flexibility*

Chinese universities exhibit the lowest reliance on all three factors, with less than 20% granting faculty decision-making authority, incorporating stakeholder feedback, or adopting adaptive policy frameworks. For example, the *Interim Measures for the Management of Generative Artificial Intelligence Services* issued by the Cyberspace Administration of China (2023) stipulates that educational use of GAI must be consistent with national security, social stability, and socialist core values, thereby placing primary authority in the hands of governmental bodies. Similarly, Ministry of Education notices emphasize cross-departmental coordination rather than local discretion. This top-down approach ensures policy uniformity and facilitates rapid implementation across institutions, but it also constrains institutional flexibility and minimizes opportunities for faculty and student participation in governance. The low prevalence of adaptive policy frameworks further suggests that GAI regulations in Chinese universities are largely fixed and predefined, with limited capacity for revision in response to technological changes.

## Discussion

### *Cross-national patterns in university GAI governance*

This study reveals distinct national patterns in how universities approach GAI governance, reflecting differences in institutional autonomy, regulatory environments, and strategic priorities. Rather than reiterating descriptive findings, these differences can be interpreted as alternative governance logics shaping policy development.

Universities in the United States tend to adopt a decentralized and innovation-oriented approach. Institutional policies emphasize faculty autonomy, experimentation, and adaptability, supported by strong engagement with frontier research and peer collaboration. This configuration enhances institutional self-efficacy and facilitates continuous policy adjustment (McDonald et al. 2024; Wu, Zhang, and Carroll 2024). However, the absence of a unified national framework may also contribute to inconsistencies in addressing ethical and regulatory risks (U.S. Department of Education, Office of Educational Technology, 2023).

In contrast, Japanese universities exhibit a more coordinated and risk-aware model, where institutional policies are closely aligned with government guidance. This approach prioritizes ethical considerations, transparency, and compliance, contributing to policy consistency and public trust (MEXT, 2023). At the same time, this centralized orientation may limit flexibility and reduce the extent to which institutions explore innovative applications of GAI.

Chinese universities demonstrate a highly centralized, state-driven governance structure. Policy development is strongly influenced by national priorities, with limited institutional autonomy in shaping guidelines (Yang 2014). While this model enables rapid alignment with national strategies and promotes large-scale technological integration, it also constrains institutional self-efficacy and limits adaptive policy development. The observed gap between technological adoption and policy formulation further highlights the challenges of balancing control and flexibility in this context (Marginson 2011).

Taken together, these patterns suggest that university GAI governance is not only shaped by technological considerations but also deeply embedded in broader institutional and socio-political structures (Mok 2005). Understanding these underlying logics is essential for interpreting cross-national differences in policy approaches.

### *Developing and applying the UPDF-GAI framework*

This study proposes the UPDF-GAI framework as a structured yet flexible tool for understanding and guiding university GAI policy development. The framework integrates five interrelated domains – PU and PEOU, PR, FC, SI, and SE – capturing technological, organizational, and social dimensions of governance.

Rather than serving as a prescriptive model, the framework provides a conceptual structure that enables institutions to interpret and evaluate their policy orientations. By linking perceived benefits and risks with institutional capacity and external influences, it offers a comprehensive perspective on how universities navigate the tensions between innovation, regulation, and ethical responsibility.

While the framework is primarily conceptual, its practical implications can be illustrated as follows.

First, institutions can map their existing policies and guidelines onto the five domains of the framework – PU and PEOU, PR, FC, SI, and SE – to identify strengths and gaps. For example, universities may assess whether their policies sufficiently address ethical risks, provide clear guidance for use, or offer adequate institutional support such as training and technical resources.

Second, institutions can prioritize specific domains based on their institutional context, regulatory environment, and strategic goals. For instance, universities operating in highly regulated environments may place greater emphasis on risk management and compliance, while those seeking to promote innovation may focus more on enhancing perceived usefulness and FC for AI adoption.

Third, institutions can iteratively refine their policies by balancing innovation and risk while strengthening institutional capacity. This may involve updating guidelines in response to technological developments, expanding support mechanisms such as training programs, and incorporating feedback from faculty and students to improve clarity and usability.

This process-oriented approach allows universities to use the framework as a diagnostic and reflective tool for policy development and evaluation, without imposing a uniform or prescriptive model across diverse institutional contexts (Figure 7).

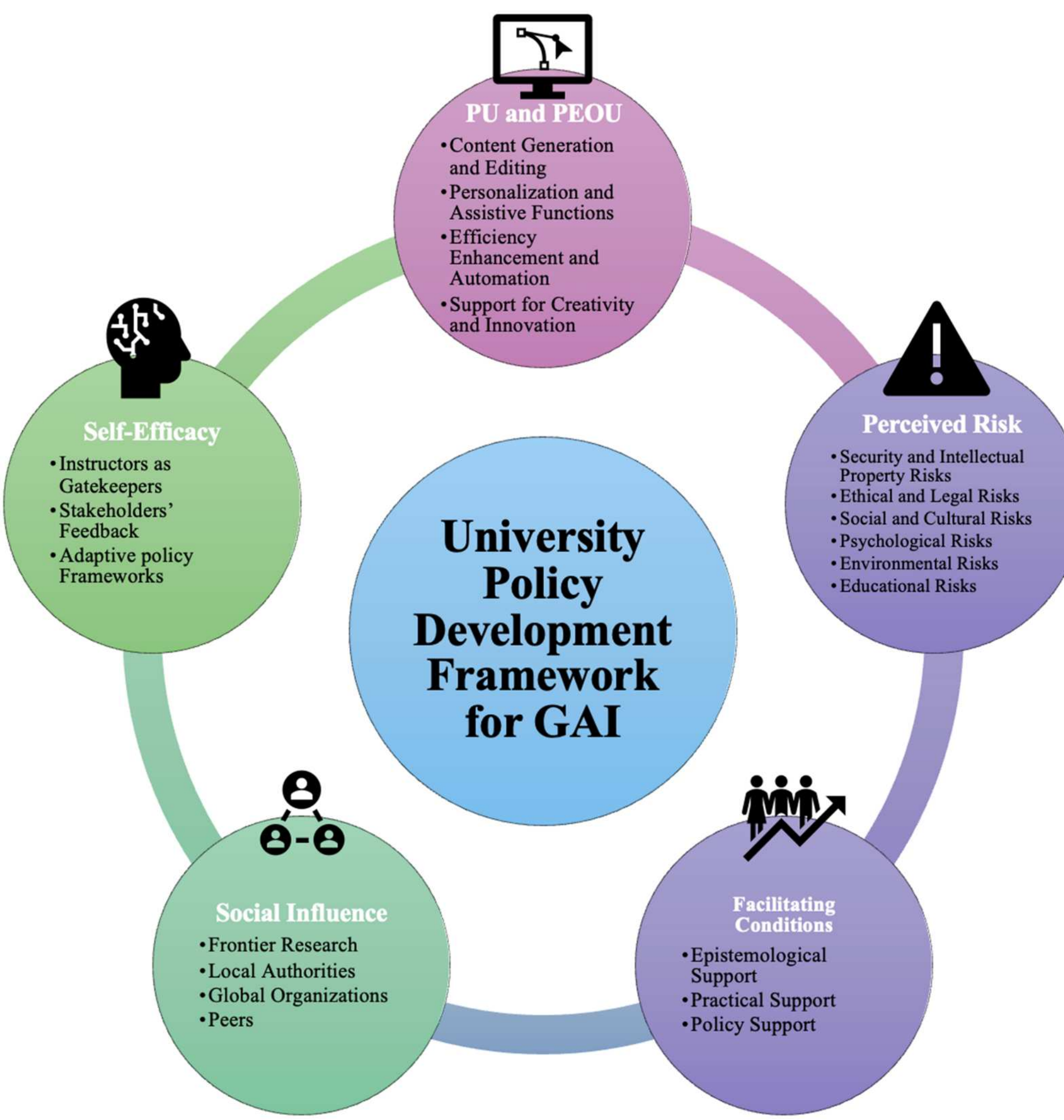

**Figure 7.** UPDF-GAI framework. The framework encompasses five key domains and their associated themes.

## Limitations

This study is limited by uneven sample distribution, particularly in China, where only six universities and seven policy documents were analyzed, which constrains the representativeness of the findings. Because the sample focuses mainly on elite universities, it may not capture the full diversity of policy practices across institutional tiers and regions. The analysis is also time-bound, as AI governance frameworks and institutional policies continue to evolve rapidly, especially in the United States and China; therefore, the findings reflect a particular stage in the development of GAI policy rather than a stable configuration. Future research could broaden institutional coverage, adopt longitudinal approaches, and integrate computational methods such as topic modeling or text-based classification to complement qualitative analysis and deepen understanding of policy trends.

## Conclusion

The comparative analysis underscores the diverse priorities and strategic orientations shaping GAI policy development across universities in different national contexts. U.S. institutions place strong emphasis on faculty autonomy, practical application, and policy adaptability – an approach deeply influenced by cutting-edge research and collaborative peer networks. In contrast, Japanese universities adopt a more government-regulated framework, foregrounding ethical considerations and risk management, yet offering limited institutional support for implementation and innovation flexibility. Meanwhile, Chinese universities pursue a centralized, state-directed model that prioritizes technological application over early-stage policy formulation, while actively advancing GAI integration into educational and research practices. Collectively, these variations highlight how cultural, institutional, and governance structures inform the trajectory of GAI policy evolution worldwide. By offering a comparative analysis of GAI guidelines among leading universities in these three influential nations, this study contributes to the emerging understanding of how higher education institutions worldwide are navigating the challenges and opportunities presented by this disruptive technology. Through multilingual analysis and the integration of findings into the UPDF-GAI framework, our work advances the conversation on institutional responses to generative AI in academia.

## Disclosure statement

No potential conflict of interest was reported by the author(s).

## Funding

This work was supported by Japan Society for the Promotion of Science KAKENHI Grant Number 24K06100.

## ORCID

*Ming Li* http://orcid.org/0000-0001-8855-1138
*Qin Xie* http://orcid.org/0009-0000-0821-8352
*Ariunaa Enkhtur* http://orcid.org/0000-0003-1544-8118
*Shuoyang Meng* http://orcid.org/0009-0002-3109-3513
*Lilan Chen* https://orcid.org/0000-0002-5367-3239
*Beverley Anne Yamamoto* http://orcid.org/0000-0003-1791-6827
*Fei Cheng* http://orcid.org/0000-0001-5161-0544
*Masayuki Murakami* http://orcid.org/0000-0003-1649-196X

## Appendix 1. Note: Links refer to datasets available as of November 30, 2024; accessibility may have changed since then.

| Top 30 U.S. News Best Colleges 2024 | 'Double First-Class' Universities in China | 'Top Global Project University' in Japan |
|---|---|---|
| Princeton University<br>https://odoc.princeton.edu/about/official-deans-communications/2023/generative-ai-teaching-and-learning-fall-2023-faculty-and<br>https://mcgraw.princeton.edu/generative-ai-and-our-classrooms<br>https://universityservices.princeton.edu/news/posted-91423-guidelines-university-data-external-artificial-intelligence-ai-solutions | Peking University | Tohoku University<br>https://olg.cds.tohoku.ac.jp/forstudents/ai-tools<br>https://olg.cds.tohoku.ac.jp/forstaff/ai-tools |
| Massachusetts Institute of Technology<br>https://ist.mit.edu/ai-guidance<br>https://ist.mit.edu/ai | Renmin University of China | The University of Tokyo<br>https://utelecon.adm.u-tokyo.ac.jp/docs/20230403-generative-ai/<br>https://utelecon.adm.u-tokyo.ac.jp/docs/ai-tools-in-classes/<br>https://utelecon.adm.u-tokyo.ac.jp/docs/ai-tools-in-classes-students/ |
| Harvard University<br>https://www.huit.harvard.edu/ai<br>https://provost.harvard.edu/guidelines-using-chatgpt-and-other-generative-ai-tools-harvard<br>https://provost.harvard.edu/guidelines-using-chatgpt-and-other-generative-ai-tools-harvard | Tsinghua University | The University of Osaka<br>https://www.tlsc.osaka-u.ac.jp/project/generative_ai/<br>https://www.osaka-u.ac.jp/ja/news/topics/2023/04/17001 |
| Stanford University<br>https://communitystandards.stanford.edu/generative-ai-policy-guidance<br>https://uit.stanford.edu/security/responsibleai<br>https://teachingcommons.stanford.edu/teaching-guides/artificial-intelligence-teaching-guide | Beihang University | Waseda University<br>https://www.waseda.jp/top/news/89507 |

Continued.

| Top 30 U.S. News Best Colleges 2024 | 'Double First-Class' Universities in China | 'Top Global Project University' in Japan |
|---|---|---|
| Yale University<br>https://provost.yale.edu/news/guidelines-use-generative-ai-tools<br>https://poorvucenter.yale.edu/ai-guidelines | Beijing Institute of Technology | Science Tokyo<br>https://www.titech.ac.jp/student/students/news/2024/069373 |
| University of Pennsylvania<br>https://isc.upenn.edu/security/AI-guidance | China Agricultural University<br>https://jszx.cau.edu.cn/art/2024/11/5/art_30715_1043316.html | University of Tsukuba<br>https://www.tsukuba.ac.jp/news/pdf/2023051116540000.pdf |
| California Institute of Technology<br>https://www.imss.caltech.edu/services/ai<br>https://ctlo.caltech.edu/universityteaching/resources/resources-for-teaching-in-the-age-of-ai | Beijing Normal University<br>https://m.thepaper.cn/newsDetail_forward_26845675 | Keio University<br>https://www.students.keio.ac.jp/com/class/registration/chatgpt.html |
| Duke University<br>https://ctl.duke.edu/ai-and-teaching-at-duke-2/<br>https://ctl.duke.edu/ai-and-teaching-at-duke-2/artificial-intelligence-policies-in-syllabi-guidelines-and-considerations/ | Minzu University of China | Hiroshima University<br>https://momiji.hiroshima-u.ac.jp/momiji-top/learning/generative_ai.html |
| Brown University<br>https://provost.brown.edu/communications/potential-impact-ai-our-academic-mission | Nankai University | Hokkaido University<br>https://www.open-ed.hokudai.ac.jp/news-2/31-3.html |
| Johns Hopkins University<br>https://engineering.jhu.edu/cmts/chatgpt/<br>https://teaching.jhu.edu/university-teaching-policies/generative-ai/ | Tianjin University | Kyushu University<br>https://www.kyushu-u.ac.jp/ja/notices/view/2482<br>https://mirai.kyushu-u.ac.jp/topics/1463/<br>https://www.kyushu-u.ac.jp/ja/notices/view/2547/ |

Continued.

| Top 30 U.S. News Best Colleges 2024 | 'Double First-Class' Universities in China | 'Top Global Project University' in Japan |
|---|---|---|
| Northwestern university<br>https://www.it.northwestern.edu/about/policies/guidance-on-the-use-of-generative-ai.html<br>https://ai.northwestern.edu/education/generative-ai-as-a-study-partner.html<br>https://ai.northwestern.edu/education/instructor-resources.html | Dalian University of Technology | Nagoya University<br>https://www.nagoya-u.ac.jp/academics/curriculum/generative_ai/index.html<br>https://www.nagoya-u.ac.jp/about-nu/declaration/ai/index.html |
| Columbia university<br>https://provost.columbia.edu/content/office-senior-vice-provost/ai-policy | Northeastern University | Tokyo Medical and Dental University<br>https://www.tmd.ac.jp/cmn/rules/houki/7hen/4shou/2setsu/74208aiguideline.pdf |
| Cornell University<br>https://it.cornell.edu/ai/ai-guidelines<br>https://teaching.cornell.edu/generative-artificial-intelligence<br>https://ai.cornell.edu/generative-ai/ | Jilin University | Kyoto University |
| University of Chicago<br>https://intranet.uchicago.edu/tools-and-resources/it-services/<br>https://genai.uchicago.edu/resources/students | Harbin Institute of Technology | Chiba University |
| University of California, Berkeley<br>https://docs.google.com/document/d/1tDz0BR6rm57iZFoVlo8RkYIUOePhMZLE7vbBAoKBbUc/edit?tab=t.0#heading=h.2vfrgujomtpx<br>https://oercs.berkeley.edu/privacy/appropriate-use-generative-ai-tools<br>https://teaching.berkeley.edu/teaching-strategies/genai-teaching/understanding-genai-and-campus-expectations | Fudan University<br>https://news.fudan.edu.cn/2024/1220/c3163a143685/page.htm | Tokyo University of Foreign Studies<br>https://www.tufs.ac.jp/documents/education/guideline/ai_guideline.pdf |

Continued.

| Top 30 U.S. News Best Colleges 2024 | 'Double First-Class' Universities in China | 'Top Global Project University' in Japan |
|---|---|---|
| University of California, Los Angeles<br>https://docs.google.com/document/d/<br>1V9OpwizUiHgLVMeEx-7g3AV1QrTAkBo0D-dsFv0DlwM/<br>edit?tab=t.0#heading= h.c6al1vj9d3gd<br>https://senate.ucla.edu/news/teaching-guidance-<br>chatgpt-and-related-ai-developments<br>https://dts.ucla.edu/initiatives/ai | Tongji University | Toyo University<br>https://www.toyo.ac.jp/news/20230727-00003.html |
| Rice University<br>https://oit.rice.edu/ai-usage-guidelines<br>https://provost.rice.edu/communications/faculty-<br>guidance-generative-ai | Shanghai Jiao Tong University<br>https://ctld.sjtu.edu.cn/storage/file/2023/12/<br>2b30798c781f304ebfc83e66faef31fe.pdf<br>https://ctld.sjtu.edu.cn/news/detail/1143 | Shibaura Institute of Technology<br>https://www.shibaura-it.ac.jp/headline/detail/<br>nid00003130.html |
| Dartmouth College<br>https://dcal.dartmouth.edu/resources/teaching-<br>methods/teaching-generative-artificial-intelligence<br>https://services.dartmouth.edu/TDClient/1806/Portal/<br>KB/ArticleDet?ID = 155837<br>https://policies.dartmouth.edu/policy/guidelines-using-<br>generative-artificial-intelligence-genai-coursework | East China Normal University | Tokyo University of the Arts |
| Vanderbilt University<br>https://hr.vanderbilt.edu/policies/generative-ai/<br>https://as.vanderbilt.edu/gci-ai/syllabus-ai-policies/<br>https://www.vanderbilt.edu/generative-ai/research/<br>https://www.vanderbilt.edu/generative-ai/faculty/<br>https://www.vanderbilt.edu/generative-ai/students/ | Nanjing University | Meiji University<br>https://www.meiji.ac.jp/gakucho/message/<br>20230517_generative-ai.html |
| University of Notre Dame<br>https://ai.nd.edu/ai-in-action/policies-and-guidelines/<br>https://honorcode.nd.edu/official-statement-regarding-<br>generative-ai-may-2023/<br>https://honorcode.nd.edu/ai-recommendations-for-<br>instructors/<br>https://learning.nd.edu/learning-technology/lab-for-ai-<br>in-teaching-and-learning-laitl/ | Southeast University | International Christian University<br>https://www.icu.ac.jp/news/2305241300.html |
| University of Michigan – Ann Arbor<br>https://genai.umich.edu/resources/students<br>https://genai.umich.edu/resources/faculty<br>https://genai.umich.edu/resources/staff | Zhejiang University<br>https://mp.weixin.qq.com/s?__biz = Mzg2ODY2ODI4OQ == &mid =<br>2247489480&idx = 1&sn = a02f3192db3926ac6de2aeaa22c748c1&chksm =<br>cea99126f9de183089a8a2639cbb67059<br>50382c559c914fee219183c0c459935ad30de6c2962&cur_album_id =<br>3036073970057543680&scene = 189#wechat_redirect | Sophia University<br>https://piloti.sophia.ac.jp/assets/uploads/2023/03/<br>23f430e7f216cbe188652f8a6855c493.pdf?<br>redirect=1<br>https://piloti.sophia.ac.jp/assets/uploads/2023/10/<br>b528436042d4a6c9dbb1e173398b3900.pdf?<br>redirect=1 |

(*Continued*)

Continued.

| Top 30 U.S. News Best Colleges 2024 | 'Double First-Class' Universities in China | 'Top Global Project University' in Japan |
| --- | --- | --- |
| Georgetown University<br>https://cndls.georgetown.edu/resources/ai/ | University of Science and Technology of China | Hosei University<br>https://info.hosei-kyoiku.jp/wp-content/uploads/2023/12/生成AIツールに対する基本的考え方（教員向け結合版）.pdf |
| University of North Carolina at Chapel Hill<br>https://www.unc.edu/search/?term = Generative ± AI ± Guideline±#gsc.tab = 0&gsc.q = Generative%20AI%20Guideline%20&gsc.page = 1<br>https://provost.unc.edu/student-generative-ai-usage-guidance/<br>https://handbook.unc.edu/generativeai.html<br>https://provost.unc.edu/generative-ai-usage-guidance-for-the-research-community/<br>https://provost.unc.edu/staff-generative-ai-usage-guide/<br>https://sph.unc.edu/iis/syllabus-guidelines-for-generative-ai/ | Xiamen University | Rikkyo University<br>https://www.rikkyo.ac.jp/news/2023/05/mknpps0000027wkb.html |
| Carnegie Mellon University<br>https://www.cmu.edu/computing/services/ai/index.html<br>https://www.cmu.edu/career/students-and-alumni/resource-library/ai-guidance/index.html | Shandong University | Soka University<br>https://www.soka.ac.jp/news/2023/10/8853/ |
| Emory University<br>https://guides.libraries.emory.edu/AI/teaching | Ocean University of China | Kanazawa University |
| University of Virginia<br>https://genai.provost.virginia.edu/<br>https://provost.virginia.edu/faculty-handbook/generative-ai-genai-teaching-and-learning | Wuhan University | Nagaoka University of Technology |

(*Continued*)

Continued.

| Top 30 U.S. News Best Colleges 2024 | 'Double First-Class' Universities in China | 'Top Global Project University' in Japan |
|---|---|---|
| Washington University in St. Louis<br>https://it.wustl.edu/ai/<br>https://ctl.wustl.edu/resources/generative-ai-and-teaching/ | Huazhong University of Science and Technology | Toyohashi University of Technology<br>https://www.tut.ac.jp/news/241128-22286.html |
| University of California, Davis<br>https://cee.ucdavis.edu/introduction-generative-ai<br>https://iet.ucdavis.edu/news/sensitive-data-and-ai-guidance<br>https://guides.library.ucdavis.edu/genai/teaching<br>https://guides.library.ucdavis.edu/genai/citing | Hunan University | International University of Japan |
| University of California, San Diego<br>https://blink.ucsd.edu/technology/ai/ai-ucsd.html<br>https://univcomms.ucsd.edu/resources/ai-guidance/index.html<br>https://academicintegrity.ucsd.edu/choose-integrity/gen-ai/index.html | Central South University | Okayama University<br>https://www.okayama-u.ac.jp/tp/news/news_id12079.html<br>https://www.okayama-u.ac.jp/tp/news/news_id12081.html |
| University of Florida<br>https://teach.ufl.edu/artificial-intelligence-in-teaching-and-learning/<br>https://privacy.ufl.edu/laws-and-regulations/ai-governance/ | National University of Defense Technology | Kumamoto University<br>https://www.kumamoto-u.ac.jp/whatsnew/jouhou/gai_gakusei<br>https://www.kumamoto-u.ac.jp/whatsnew/jouhou/gai_kyosyokuin |
| University of Southern California<br>https://libguides.usc.edu/generative-AI/introduction<br>https://libguides.usc.edu/generative-AI/usc-guidelines<br>https://academicsenate.usc.edu/wp-content/uploads/sites/6/2023/02/CIS-Generative-AI-Guidelines-20230214.pdf#:~:text = URL%3A%20https%3A%2F%2Facademicsenate.usc.edu%2Fwp | Sun Yat-sen University | Ritsumeikan Asia Pacific University<br>https://www.apu.ac.jp/home/Usage_of_Generative_AI_at_APU_J.pdf |
| | South China University of Technology | Akita International University |

(*Continued*)

Continued.

| Top 30 U.S. News Best Colleges 2024 | 'Double First-Class' Universities in China | 'Top Global Project University' in Japan |
| --- | --- | --- |
| | Sichuan University | University of Aizu |
| | https://lib.scu.edu.cn/genai/home.html | https://u-aizu.ac.jp/files/cd87075ead451e35bcfebed5b79f71c88450a7d1.pdf |
| | University of Electronic Science and Technology of China | Kyoto Institute of Technology |
| | Chongqing University | Nara Institute of Science and Technology |
| | Xi'an Jiaotong University | Ritsumeikan University |
| | | https://www.ritsumei.ac.jp/news/detail/?id=3153 |
| | Northwestern Polytechnical University | Kansei Gakuin University |
| | Northwest A&F University | |
| | Lanzhou University | |
| | Zhengzhou University | |
| | Xinjiang University | |
| | Yunnan University | |